\documentclass[pre,twocolumn,amsmath,amssymb,aps,showpacs]{revtex4}

\usepackage{graphicx}

\begin{document}

\title{Bistable phase control via rocking in a nonlinear electronic oscillator}

\author{Javier M. Buld\'u$^{1,2}$, K. Staliunas$^{1,3}$, J.A. Casals$^1$,
Jordi Garcia-Ojalvo$^1$}

\affiliation{
$^1$Departament de F\'{\i}sica i Enginyeria Nuclear, Universitat
Polit\`ecnica de Catalunya, Colom 11, 08222 Terrassa, Spain.\\
$^2$Nonlinear Dynamics and Chaos Group, 
Departamento de F\'{\i}sica Aplicada y
Ciencias de la Naturaleza, Universidad Rey Juan Carlos, Tulip\'an s/n,
28933 M\'ostoles, Madrid, Spain.\\
$^3$Instituci\'o Catalana de Recerca i Estudis Avan\c{c}ats (ICREA),
Colom 11, E-08222 Terrassa, Barcelona, Spain.
}

\begin{abstract}
We experimentally demonstrate the effective rocking of a nonlinear electronic circuit
operating in a periodic regime. Namely, we show that driving a Chua circuit with
a periodic signal, whose phase alternates (also periodically) in time, we lock the oscillation frequency of the circuit to that of the driving signal, and its phase to 
one of two possible values shifted by $\pi$, and lying between the alternating phases of the input signal. 
In this way, we show that a rocked
nonlinear oscillator displays phase bistability. We interpret the experimental results
via a theoretical analysis of rocking on a simple oscillator model, based on a
normal form description (complex Landau equation) of the rocked Hopf bifurcation. 
\end{abstract}

\maketitle

\section{Introduction}

Multistability is a very useful property of nonlinear systems, which lends itself
to relevant applications such as memory storage and pattern recognition
\cite{foss96}. An intriguing version of the phenomenon that can take place in
nonlinear oscillators
is that of phase bistability, whereby the phase of the oscillator can lock to one of
two possible values for a given set of operating conditions. Since the two states differ only in phase but not in amplitude, phase bistability provides a way of storing digital
information without involving abrupt changes in average energies.

The phase of most nonlinear oscillators is invariant, i.e. the system
is indifferent to arbitrary variations of the phase. A standard way of forcing a
nonlinear oscillator to have a fixed phase is by phase locking. This procedure consists
simply on forcing the nonlinear oscillator with a harmonic signal whose frequency
is close to its natural frequency. In those conditions, the system
may lock its frequency and phase to that of the external driving. A constant
phase difference arises between the injected signal and the system output, which
depends on the frequency mismatch, injection strength and internal parameters of the system \cite{pik00}. Obviously, phase
locking leads to a monostable phase response of the system.

Recently, a phase locking procedure was proposed that leads to a bistable phase
response \cite{sta03}. The method, called {\em rocking}, relies on subjecting
the driving (harmonic) signal to an additional modulation (e.g. a sinusoidal modulation
of its amplitude). In this situation, the phase of
the output oscillations can still lock to that of the input signal, but (somewhat counterintuitively) with a $\pm\pi/2$-shift relative to the input phase.
In other words, the oscillator locks its phase to one of two values shifted by $\pi$,
which results in a bistable response: for a given set of parameter values
(including the parameters of the driving signal), the system can be forced
to oscillate in one of two different regimes depending on the initial condition.

Rocking was initially proposed in the context of broad-area lasers \cite{sta03},
where transverse patterns are
notoriously hampered by the lack of phase bistability, in contrast with other
well-known optical systems. In this context, rocking has been experimentally
shown to lead to stable patterns in lasers \cite{tar05}.
This is the only experimental demonstration so far of the rocking phenomenon.
However, rocking is a general physical phenomenon, and should exist in any
self-oscillating system \cite{val05}. Here we demonstrate 
for the first time the rocking effect in a nonlinear electronic oscillator.
Additionally, in contrast with the initial theoretical proposal \cite{sta03}, where
the secondary modulation of the driving field was sinusoidal (resulting in a
bichromatic master signal), here we use a periodic alternation of the phase of the
input signal, keeping its amplitude constant. Our results show that
rocking can be efficiently attained in an electronic system under these conditions. 

The paper is organized as follows. Experiments are described in
Sec.~\ref{exp_results}. Section~\ref{theory} contains a theoretical analysis
in terms of the normal form of the Hopf bifurcation, which takes the form of a
complex Landau equation (CLE) with periodic injection.

\section{Experimental observations}
\label{exp_results}

Our nonlinear electronic oscillator is the so-called Chua circuit, which is a simple
electronic circuit widely used for the generation of low
dimensional chaotic regimes \cite{mad93}. In this article we investigate the behavior
of the Chua circuit operating in a periodic regime. The oscillator is
driven by a periodic external
signal (with frequency close to the characteristic frequency of the circuit) whose
phase alternates periodically in time.

\begin{figure}[htb]
\centering
\includegraphics[width=85mm,clip]{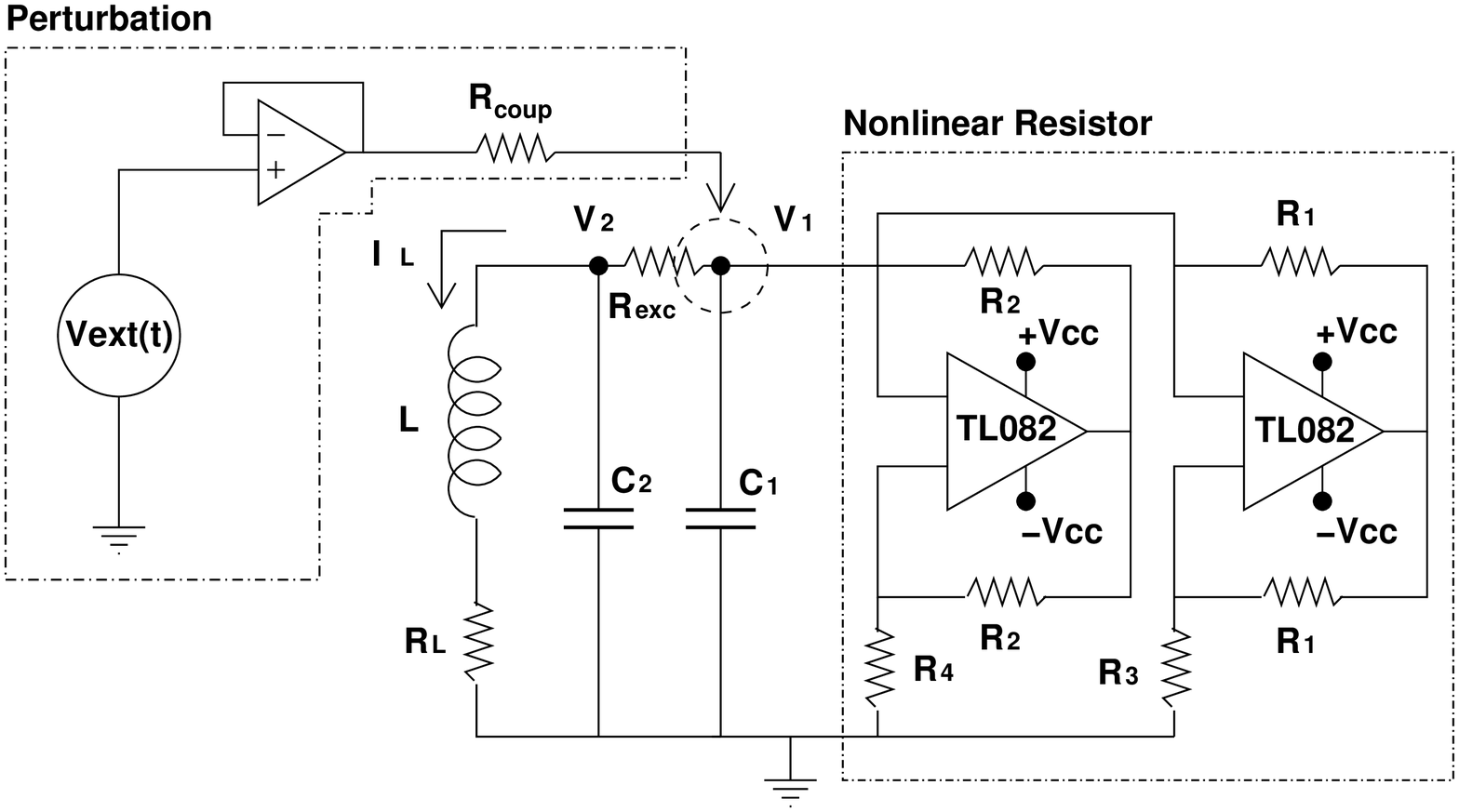}
\centering
\caption[2]{
Scheme of the electronic oscillator used in this paper. The circuit is built with two
TL082 operational amplifiers, and with passive
electronic components of values: $C_1=10$~nF, $C_2=100$~nF, $L=8$~mH,
$R_1=10\; k\Omega$,  $R_2=270\;\Omega$,
$R_3=1\; k\Omega$,  $R_4=220\;\Omega$, $R_L=20\;\Omega$. Operational
amplifiers are fed by a DC source of $V_{cc}=\pm 9.0$~V. 
We set $R_{\rm exc}=1.957\;k\Omega$ in order
to have periodic dynamics when the circuit is isolated.
The coupling resistance is $R_{\rm coup}=22\;k\Omega$.
$V_1$ and $V_2$ correspond to the outputs of the circuit.
} 
\label{fig:f01}
\end{figure}

\begin{figure}[htb]
\centering
\includegraphics[width=85mm,clip]{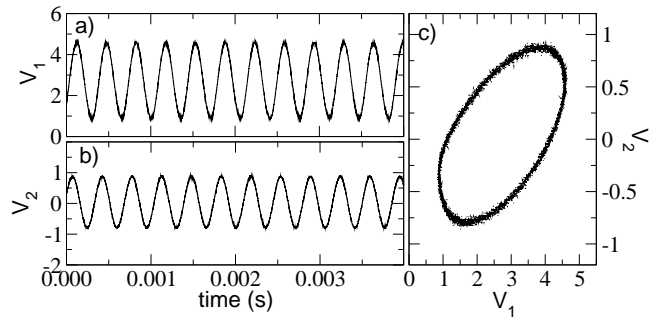}
\centering
\caption[2]{
Dynamics of the nonlinear circuit in the absence of external input. Time traces
of $V_1$ (a) and $V_2$ (b), and the corresponding phase space plot (c).
} 
\label{fig:f02}
\end{figure}

Figure \ref{fig:f01} shows a detailed description of the Chua circuit used in this paper.
A nonlinear resistor is connected to a set of passive electronic components.
We have systematically studied the dynamical ranges of the Chua circuit when 
$R_{\rm exc}$ is varied, observing stable, periodic, excitable and chaotic dynamics. 
We fix the circuit to have a periodic output by setting 
$R_{\rm exc}=1.957$~k$\Omega$. Under these conditions the 
dynamics of the circuit lies on a 
limit cycle with natural frequency $f_0=2840.5$~Hz (see Fig.~\ref{fig:f02}).
The build-up time of these oscillations (time interval necessary for the oscillation amplitude to grow a factor $e$ starting from the off state) is, for the
experimental parameters chosen, around 15~ms.

We use an external voltage source (of amplitude $V_{\rm ext}$) in order 
to perturb the circuit. External 
input is introduced into the circuit through a voltage follower and a 
coupling resistance $R_{\rm coup}$,
which can be tuned in order to modify the coupling strength.

The dynamics of the circuit is described by the equations:
\begin{subequations}
\begin{eqnarray}
& &C_1\frac{dV_1}{dt} =
\frac{V_2-V_1}{R_{\rm exc}}-g(V_1,V_{cc})+\frac{V_{\rm ext}-V_1}{R_{\rm coup}}
\label{eq:chua_eq1}
\\
& &C_2\frac{dV_2}{dt} =
\frac{V_1-V_2}{R_{\rm exc}}+I_L
\label{eq:chua_eq2}
\\
& &L\frac{dI_L}{dt} =
-V_2-R_L I_L\,.
\label{eq:chua_eq3}
\end{eqnarray}
\label{eq:chuatot}
\end{subequations}
Hence the phase space of this system is three-dimensional.
$g(V_1,V_{cc})$ is a piecewise-linear function given by:
\begin{eqnarray}
& & g(V_1,V_{cc})=G_a V_1\nonumber \\
& & \qquad+ 0.5 (G_a-G_b)(|V_1+B_p|-|V_1-B_p|)\,.
\label{eq:piece}
\end{eqnarray}
Here $G_a$, $G_b$ are the slopes and $\pm B_p$ denote a break point \cite{ken93}.

\begin{figure}[htb]
\centering
\includegraphics[width=70mm,clip]{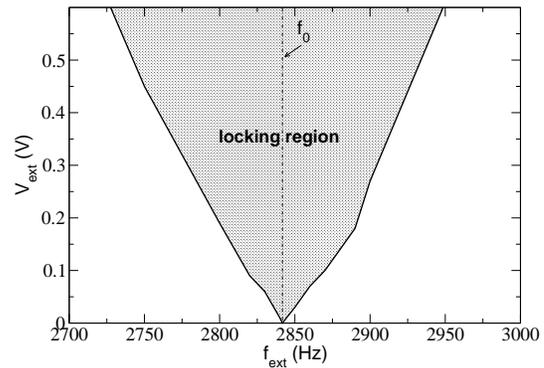}
\centering
\caption[2]{
Response of the system to an external periodic input. $f_0=2840.5$~Hz corresponds to 
the natural frequency of the circuit.
} 
\label{fig:f03}
\end{figure}

The phase of the limit cycle oscillations shown in Fig.~\ref{fig:f02} is not
defined. It can be fixed by applying to the system a periodic
signal of frequency $f_{\rm ext}$ and amplitude $V_{\rm ext}$. Figure~\ref{fig:f03}
shows the locking region of the system in the parameter space defined by
the amplitude and frequency
of the external forcing. As expected,
a resonance exists at the natural frequency of the circuit, $f_0=2840.5$~Hz.

\begin{figure}[htb]
\centering
\includegraphics[width=70mm,clip]{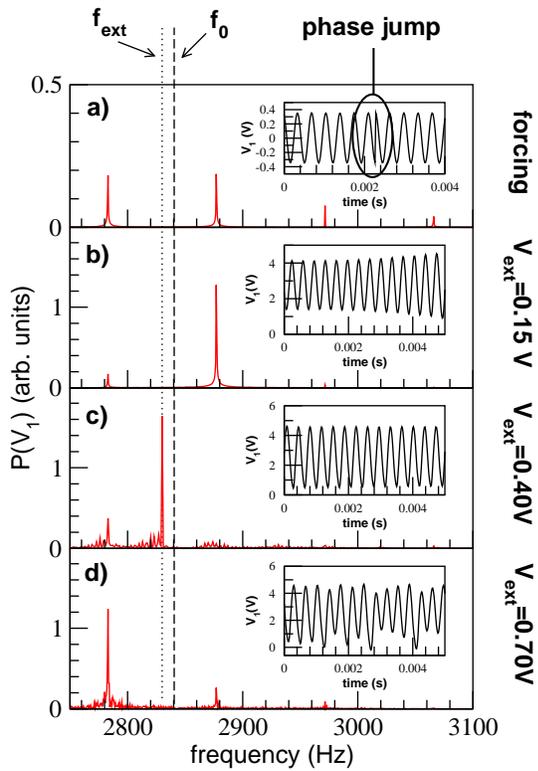}
\centering
\caption[2]{
Response of the circuit to a periodic signal
of frequency $f_{\rm ext}=2830$~Hz, amplitude $V_{\rm ext}$ (variable), and jumps
of $\pi$ in the phase with a frequency $f_{\rm jump}=94.33$~Hz.
In (a), we plot the power spectrum and the temporal evolution
(inset) of the external signal. (b), (c) and (d) show
the power spectra and the time series (insets) for
$V_{\rm ext}=0.15$~V, $V_{\rm ext}=0.40$~V and $V_{\rm ext}=0.70$~V
respectively.  
} 
\label{fig:f04}
\end{figure}

The single-frequency forcing described in the previous paragraph locks the
phase to a {\em unique} value with a well-defined shift with respect to the input
signal. Nevertheless, as mentioned above, a bistable phase response can be obtained by {\em rocking}
the oscillator. To that end, we perturb the harmonic
forcing by introducing periodic $\pi$-jumps in the phase, with frequency
$f_{\rm jump}$. Figure~\ref{fig:f04}(a) displays the power
spectrum of the input signal, which 
differs from a pure sinusoidal perturbation in the fact that now it
exhibits two maxima at frequencies
$f_{\rm ext}^{(1)}$ and $f_{\rm ext}^{(2)}$, centered around a 
frequency $f_{\rm ext}$ and separated by an amount that we denote $f_{\rm jump}$.
In the particular experimental conditions of Fig. \ref{fig:f04}, $f_{\rm ext}=2830$~Hz and
$f_{\rm jump}=94.33$~Hz. The inset shows the corresponding time
series, where a jump of $\pi$ in the frequency can be observed.
Plots \ref{fig:f04}(b,c,d) show the response of the circuit for 
low, intermediate and high amplitudes of the external input, respectively.
Specifically, we can observe an entrainment to $f_{\rm ext}^{(1)}$ for low
couplings, a shift to $f_{\rm ext}$ for intermediate couplings (note 
that this frequency is not present at spectrum of the external input), and finally, an 
entrainment to $f_{\rm ext}^{(2)}$ when the coupling is high enough.   

The situation depicted in Fig.~\ref{fig:f04}(c) constitutes an example of rocking:
the system is perturbed with a periodic signal with phase 
jumps of $\pi$ (alternating with frequency $f_{\rm jump}$),
and as a result an output phase arises that is different from any of the two
alternating inputs, but equal to their mean value. In other words,
the system is not able to follow the phase jumps due to their relatively high frequency,
and adopts a phase located between the two alternating values of the input
phase. The rocking effect can be observed by comparing the spectrum
of the input signal with that of the circuit. When the circuit
is entrained to the frequency $f_{\rm ext}$, not present at the input spectrum,
we have rocking [compare Figs. \ref{fig:f04}(a) and (c)].
 
\begin{figure}[htb]
\centering
\includegraphics[width=80mm,clip]{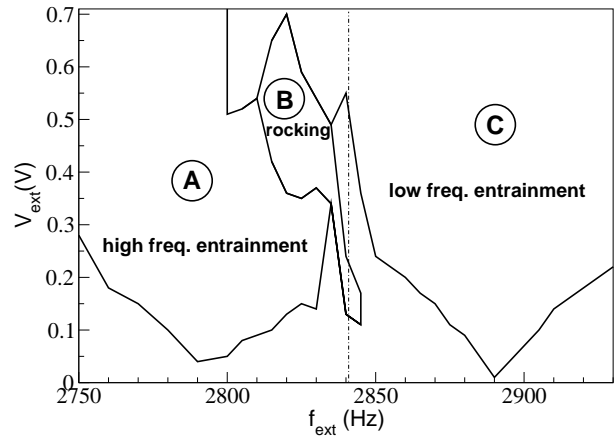}
\centering
\caption[2]{
Entrainment regimes in the parameter space of $f_{\rm ext}$ and $V_{\rm ext}$ for 
a given $f_{\rm jump}=94.33$~Hz. The dashed line corresponds to the natural
frequency of the circuit ($f_0=2840.50$~Hz). In the region labeled A, the circuit is entrained to $f_{\rm ext}^{(2)}=f_{\rm ext}+f_{\rm jump}/2$.
Rocking is observed at region B. In region C the system
is entrained to the lower frequency of the input spectrum, 
$f_{\rm ext}^{(1)}=f_{\rm ext}-f_{\rm jump}/2$.
} 
\label{fig:f05}
\end{figure}

We now examine the dependence of the rocking phenomenon on the amplitude
and frequency of the input signal. Figure \ref{fig:f05} shows
the rocking region in the parameter plane defined by those two quantities, which
is closed and surrounded by other two regions where
the system is entrained to either $f_{\rm ext}^{(1)}$ or $f_{\rm ext}^{(2)}$, 
as we have already seen in Figs.~\ref{fig:f04}(b) and (d).
This result can be explained by taking into account the spectrum of the 
input signal, with two peaks ($f_{\rm ext}^{(1)}=f_{\rm ext}-f_{\rm jump}/2$ 
and $f_{\rm ext}^{(2)}=f_{\rm ext}+f_{\rm jump}/2$),
and the locking region given by Fig.~\ref{fig:f03}. 
Let us consider first a rocked input signal with a low mean frequency
$f_{\rm ext}$
lying outside (on the left of) the locking region. 
When shifting the spectrum of the input signal (by modifying $f_{\rm ext}$) 
from low to high frequencies, the peak at
the highest frequency ($f_{\rm ext}^{(2)}$) is the first to enter the
locking region, entraining the system at $f_{\rm ext}^{(2)}$. This entrainment
has a resonance at $f_{\rm ext}^{(2)}=f_0$ (or $f_{\rm ext}=f_0-f_{\rm jump}/2$),
which corresponds to the 
minimum observed at region A of Fig.~\ref{fig:f05}. 

A similar situation
occurs when shifting the input signal from high to low frequencies
from an starting point placed outside (now, at the right of) the locking region. In this
case, the peak corresponding to the lower frequency $f_{\rm ext}^{(1)}$ enters first
the locking region, and entrains the system to $f_{\rm ext}^{(1)}$ (region C in
Fig.~\ref{fig:f05}). As in the previous
case, there is a resonance at $f_{\rm ext}^{(1)}=f_0$ (or $f_{\rm ext}=f_0+f_{\rm jump}/2$).
We find rocking between the two regions of entrainment, where the system
is pulled from both sides and none of them has enough influence to overcome
the other. The rocking region (B in Fig. \ref{fig:f05}) also shows a resonance
appearing, as it would be expected, at $f_{\rm ext}=f_0$.
 
It is worth noting that the rocking region is closed
in ($f_{\rm ext}$,$V_{\rm ext}$) space, which means that not
only frequency but also amplitude must be properly tuned. Indeed: for
low amplitudes, the system
can not be entrained and oscillates with its
natural frequency. On the other hand, when amplitude is too high, the system
follows the phase jumps of the input frequency, i.e. $f_{\rm ext}^{(1)}$ and 
$f_{\rm ext}^{(2)}$, and rocking is not observed.
Only for moderate amounts of coupling the system is perturbed
by the input signal, but not enough to follow it, laying at a 
frequency between $f_{\rm ext}^{(1)}$ and $f_{\rm ext}^{(2)}$. This is what we
call rocking. 

\begin{figure}[htb]
\centering
\includegraphics[width=90mm,clip]{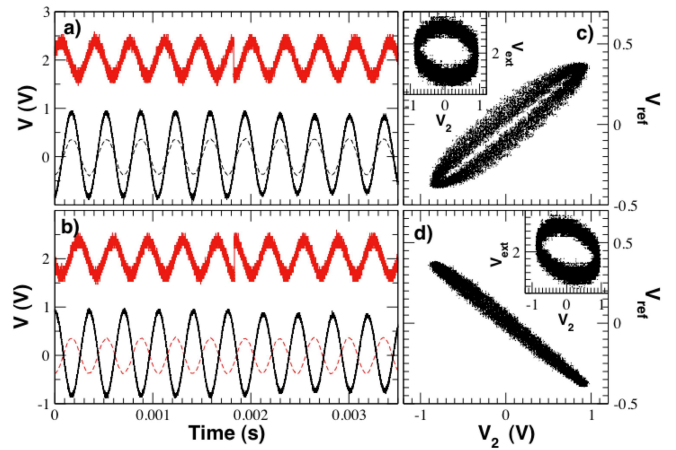}
\centering
\caption[2]{
Phase bistability of the rocking state. (a,b) Time series of the external perturbation
$V_{\rm ext}$ (top trace), the response of the system $V_2$ (solid bottom trace), and a
reference sinusoidal signal $V_{\rm ref}$ (solid dashed trace). (c,d) Synchronization plots
in the phase plane ($V_{2},V_{\rm ref}$); the insets show the corresponding plot
in the plane ($V_{2},V_{\rm ext}$).
} 
\label{fig:f07}
\end{figure}

So far we have identified the rocked state in terms of the frequency response
of the system. But as mentioned in the Introduction, and as described in the
next section, two different solutions, with a phase difference of $\pi$, are possible
when the system is rocked. Taking the phase of the rocked
system as a reference $\phi_{\rm rock}$, the input signal, which consists of a sine
wave with phase jumps of $\pi$, has a phase
$\phi_{\rm ext}=\phi_{\rm rock}+\phi_0 \pm \frac{\pi}{2}$, where $\phi_0$ is a certain phase difference due to the impedance of the circuit 
and $\pm \frac{\pi}{2}$ accounts for the phase jumps of the external signal, which
occur with a frequency $f_{\rm jump}$. When $\phi_0\sim0$, the phase difference
between the input and the output $\phi_{\rm ext}-\phi_{\rm rock}$ is
$\pm \frac{\pi}{2}$. In order to distinguish experimentally between the two possible 
solutions of the rocked state, we proceed in the following way: starting with the
circuit operating under the influence of the rocking signal, we define a reference
sinusoidal signal whose frequency and phase are matched with that of the
rocked output state. We then
reduce the amplitude of the external signal with the aim of losing the rocking
of the system. When rocking is lost, we increase again the amplitude of the external
perturbation up to its initial value, in order to recover the rocking, and compare 
the reference sinusoidal 
signal with the output of the system. A possible result is shown in Fig.~\ref{fig:f07},
in which (a) plots the external perturbation $V_{\rm ext}$ (in red), the reference signal 
$V_{\rm ref}$ (black dashed line),
and the response of the circuit $V_2$ (black line). It can be observed that 
the output voltage is in phase with the reference signal, while it has a
$\pm\frac{\pi}{2}$ phase difference with the input signal [compare the phase-plane
plot in Fig.~\ref{fig:f07}(c) with that in its inset].
It can also happen that, upon reentrance in the rocking region, the situation
resembles instead that in plots (b) and (d) of Fig.~\ref{fig:f07}.
In that case, while the output
signal keeps a $\pm\frac{\pi}{2}$ phase difference with the external perturbation
[see inset in plot (d)],
it is however in antiphase (i.e. has a phase difference of $\pi$)
with the reference signal [see main graph in plot (d) of Fig.~\ref{fig:f07}].
This fact reveals that the circuit is now operating at the other solution,
which is shifted $\pi$ with respect with the previous one.   

Both solutions are possible and stable when the system is inside
the rocking region, choice of one or another depends on the initial conditions.
In spite of their stability, spontaneous jumps between the two solutions, driven by 
internal noise in the circuit, is possible near the border of the rocking region.
The effect of noise was not investigated in detail in the present article. Further
work in that direction is in progress.

\section{The normal form description}
\label{theory}

The nonlinear circuit used in the experiments described above is a self-oscillatory
system, where oscillation sets in after a Hopf bifurcation \cite{moi99}.
Since we do not focus here on other dynamical regimes of the Chua circuit
(such as bistable or chaotic regimes), then the full description
[Eqs.~(\ref{eq:chua_eq1})-(\ref{eq:chua_eq3})] is not necessary in order to
understand the basic features of the phenomena reported here, and we use a theoretical description
as simple as possible. This simplicity of description allows us to obtain analytical
results, as well as an insight into the rocking process. 
The evolution of the slowly varying envelope of the oscillatory system above
(but close to) the Hopf bifurcation is in a most simple way described by the Landau equation:
\begin{eqnarray}
& & \frac{dA}{d\tau}=A+i\Delta \omega A - |A|^2 A + A_{\rm inj}(\tau),
\label{eq:landau}
\end{eqnarray}
here $\tau$ is a dimensionless time, normalized to the build-up time of the oscillations $\tau_0$, which depends on the distance to the Hopf bifurcation point. In the experimental conditions described above, it is $\tau_0=15$~ms.
The variable $A(\tau)$ denotes the slowly varying complex envelope of the variable (voltage), i.e.
$V(\tau)=A(\tau)e^{i\omega_0 t}+c.c.$
with $\omega_0$ being a reference frequency (chosen to coincide with the
injection frequency). $A_{\rm inj}(\tau)$ is the envelope of the injection signal. It is
generally a complex quantity, but here we will consider it, without the loss of
generality, as a real-valued function. $\Delta \omega$ in Eq. (\ref{eq:landau}) 
represents the mismatch between the oscillation frequency and the injection
frequency. 
Equation~(\ref{eq:landau}) can also be derived systematically as the amplitude
equation corresponding to the circuit model (\ref{eq:chua_eq1})-(\ref{eq:chua_eq3}).
To that end one can assume periodic, but slowly modulated oscillations
(providing a
slow temporal scale), then make a perturbative expansion and analyze the solvability
of the different orders of approximation.
At a certain order Eq.~(\ref{eq:landau}) appears.   

\begin{figure}[htb]
\centering
\includegraphics[width=90mm,clip]{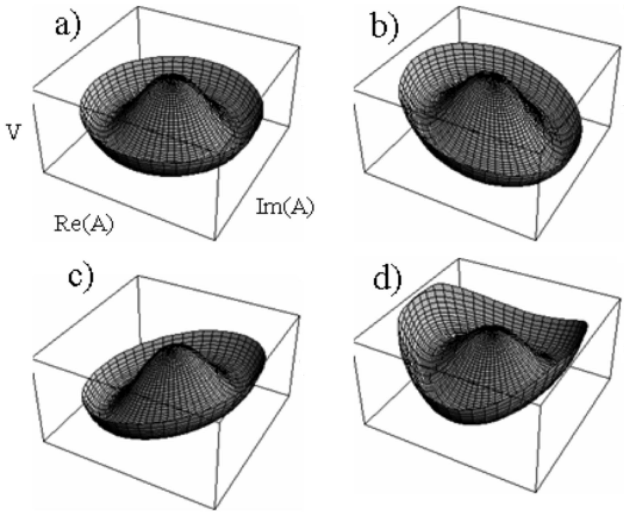}
\centering
\caption[2]{
Qualitative 3D plot of the potential associated with Eq. (\ref{eq:landau}), which
describes a rocked oscillator (arbitrary units are used). a) Without injection
($A_{\rm inj}=0$) the potential is radially symmetric, in agreement with the phase
invariance of the free running oscillator. b) With constant injection ($\Omega=0$) the
potential
tilts along the direction $\mathrm{Re}(A)$ proportionally to the forcing amplitude $F$,
and a single isolated minimum appears, corresponding to the phase-locked state
of the oscillator with injected signal; here $A_{\rm inj}<0$. c) Like (b), but with
 $A_{\rm inj}>0$. Under
rocking ($\Omega\neq 0$) the potential oscillates back and forth between the two
cases b and c, through a. Under such a forcing, the system would tend to remain
close to the imaginary axis $\mathrm{Re}(A) = 0$, around either of the two regions separated
by the local maximum around the origin. d) Effective potential associated with
Eq.~(\ref{eq:pot2}),
which is the initial potential in (a), deformed due to fast rocking.
} 
\label{fig:f03_theo}
\end{figure}

In the resonant case ($\Delta\omega=0$), eq.~(\ref{eq:landau}) is variational, since
it can be written as $dA/d\tau=-\delta V /\delta A^*$, with a potential
\begin{equation}
V=-|A|^2 +\frac{1}{2}|A|^4-\mathrm{Re}(A_{\rm inj}A^*)\,.
\label{eq:pot1}
\end{equation}
In the absence of injection $A_{\rm inj}=0$, the potential has 
the shape of a sombrero and displays a degenerated minimum along the
circumference 
$|A|^2=1$ in the complex plane $\mathrm{Re} (A)$-$\mathrm{Im} (A)$, as shown
in Fig.~\ref{fig:f03_theo}(a). 
In that case the phase of the complex envelope, $\varphi=\arg(A)$, is 
thus arbitrary. Alternatively, one says that Eq.~(\ref{eq:landau}) displays 
phase invariance.

In the presence of a constant resonant injection the potential is tilted along 
the Re$(A)$-axis, where it exhibits an isolated minimum. This is plotted in
Figs.~\ref{fig:f03_theo}(b,c). Phase invariance is thus broken,
and the phase of the output is locked to that of the
input ($\varphi=0$). In the case of an external signal with constant amplitude
but periodically alternating phase, 
$A_{\rm inj}=A_0 \;\mathrm{sign}[\sin(\Omega \tau)]$, the 
potential is periodically tilted, or rocked, around the axis $\mathrm{Im}(A)$ (hence the 
name ``rocking" \cite{sta03}). Under this rocking the system 
avoids active areas in the phase space
$\mathrm{Re}(A)$-$\mathrm{Im}(A)$
(located on the axis $\mathrm{Re}(A)$ in this case) 
and drifts to quiet areas (located along the axis $\mathrm{Im}(A)$) \cite{sta03}.
%, following the general principles of physics \footnote{Following the general idea 
%that systems avoid noisy areas of the phase space:
%See, e.g., Landau, L.D. and Lifschitz E.M., {\it Course 
%of Thoeretical Physics}, vol. 1,
%Pergamon, London, 1959. }.
The phase symmetry 
is again broken but now, unlike the usual case of constant forcing, 
the phase of the oscillations can lock to one of two values, symmetric with respect 
to the input phase, differing by $\pi$. In other words the radially symmetric 
Hopf bifurcation describing the oscillations of the autonomous system 
[of which Eq. (\ref{eq:landau}) with $A_{\rm inj}=0$ represents its normal form]
deforms into a pitchfork bifurcation due to the rocking of the potential. 

Previous studies \cite{sta03} considered a harmonically oscillating input
$A_{\rm inj}=A_0 \sin(\Omega \tau)$. Following the experiments described above,
we investigate here instead the case of a step-like modulation $A_{\rm inj}=A_0\; \mathrm{sign}[\sin(\Omega \tau)]$.
In the analytical study we consider the limit of ``strong and fast" rocking
$A_0=\epsilon^{-1} f$, $\Omega=\epsilon^{-1} \omega$, ($0<\epsilon<<1$),
and keep the rest of parameters as $O(1)$ quantities. This allows us to separate
the slow time scale $\tau$ 
of the unforced system and the fast time scale $T=\epsilon^{-1}\tau$ of rocking:
$\partial_\tau\rightarrow\epsilon^{-1}\partial_T+\partial_\tau$. 
We seek for a solution to Eq~(\ref{eq:landau}) of the form
$A(\tau)=a(\tau)+\xi(T)+{\cal O}(\epsilon)$, 
which means that we separate the order parameter into a fast 
oscillating part $\xi(T)$ (due to the rocking) and a slowly varying part $a(T)$. 
At order $\epsilon^{-1}$ we find:
\begin{eqnarray}
& & \frac{d\xi}{dT}=A_{\rm inj}(T),
\label{eq:xi}
\end{eqnarray}
leading to
\begin{eqnarray}
& & \xi(T)=\int{A_{\rm inj}(T)dT}+\mathrm{const},
\label{eq:xi_int}
\end{eqnarray} 
Assuming that the choice of integration constant results in
$\langle \xi \rangle=0$, the integration yields $\langle |\xi|^2 \xi \rangle=0$, and
$\langle |\xi|^2 \rangle=\langle \xi^2 \rangle=\gamma$, with 
$\gamma=(A_0/\Omega)^2 \pi^2 /12\approx (A_0/\Omega)^2 0.82247$.
Next we integrate Eq.~(\ref{eq:landau}) [with the {\em Ansatz} for $A(\tau)$
given above]
over the forcing period, and consider that the slow amplitude does not 
change sensibly over one period: 
\begin{eqnarray}
& & \frac{da}{d\tau}=a-(|a|^2a+a^*\langle \xi^2\rangle+2a\langle |\xi|^2\rangle)+i\Delta\omega a\,,
\label{eq:order_zero}
\end{eqnarray}
where we have used the fact that the averages of odd powers of $\xi$ are zero,
as mentioned above,  
$\langle\xi\rangle=0$, $\langle|\xi|^2\xi\rangle=0$.
Inserting the calculated averages of the even powers of $\xi$ yields: 
\begin{equation}
\frac{da}{dt}=(1-2\gamma)a+\gamma a^* + |a|^2 a +i\Delta\omega a\,.
\label{eq:order_zero_b}
\end{equation}   	
This expression differs from that in \cite{sta03} only in the coefficient $\gamma$,
equal to $\gamma=\frac{1}{2}(A_0/\Omega)^2$ in that case, where the input
signal was harmonic.

In the fully resonant limit $\Delta \omega=0$, Eq.~(\ref{eq:order_zero_b}) is
variational, and its potential reads
\begin{equation}
V=-(1-2\gamma)|a|^2 + \frac{1}{2}|a|^4 - \gamma \frac{1}{2}\left(a^2 + a^{*2}\right)\,.
\label{eq:pot2}
\end{equation}
A comparison with the potential (\ref{eq:pot1}) of Eq.~(\ref{eq:landau})
reveals that the effect of 
rocking consists in deforming the rotationally symmetric potential of the undriven 
system, so that now local minima appear at  
$a=\pm i\sqrt{1-\gamma}$, corresponding to
two symmetric points located at the axis $\mathrm{Im}(a)$.
These minima correspond to the
steady solutions of Eq.~(\ref{eq:order_zero_b}) in the resonant case. The deformation
of the potential is compatible with the general theory of motion in rapidly oscillating
fields \cite{lan59}.
 
\begin{figure}[htb]
\centering
\includegraphics[width=60mm,clip]{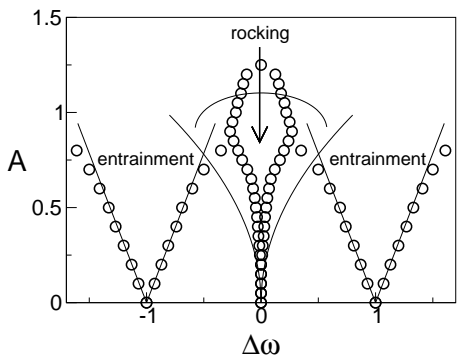}
\centering
\caption[2]{
Rocking range of a perturbed oscillator, as obtained theoretically
from Eq.~(\protect\ref{eq:range_rocking}) and (\ref{eq:lrstep}) (solid lines) and
by numerical integration of Eq. (\ref{eq:landau}) (white circles), for a rocking
frequency $\Omega= 1$.
} 
\label{fig:thlock}
\end{figure}

\begin{figure}[htb]
\centering
\includegraphics[width=60mm,clip]{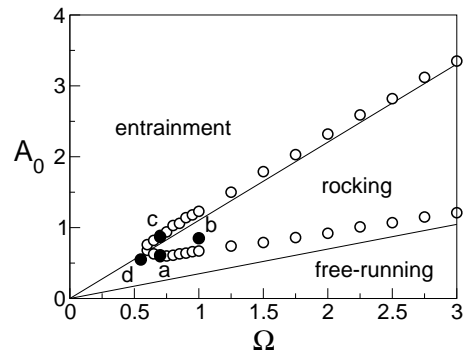}
\centering
\caption[2]{
Rocking range of a perturbed oscillator, as obtained theoretically
from Eq.~(\protect\ref{eq:range_rocking}) (solid lines) and
by numerical integration of Eq. (\ref{eq:landau}) (white circles), for a detuning
$\Delta \omega= 0.1$. Black circles denote points whose trajectories are
shown in Fig.~\protect\ref{fig:f05_theo} below.
In the entrainment region the oscillator follows adiabatically the input signal, while
in the rocking region the phase is set to the solutions obtained in
Eq.~(\ref{eq:solutions}).
} 
\label{fig:f04_theo}
\end{figure}

Equation (\ref{eq:order_zero_b}) has two relevant spatially homogeneous 
solutions (apart from the trivial one $a=0$): 
\begin{eqnarray}
& & a=\pm u \exp(i\phi)\nonumber\\
& & u^2=1-2\gamma + \sqrt{\gamma^2- \Delta \omega^2}\label{eq:solutions}\\ 
& & \phi=-\frac{1}{2}\arcsin(\Delta \omega/\gamma), 
\nonumber
\end{eqnarray} 
of equal intensities but opposite phases. The existence range of these ``rocked states"
is 
\begin{eqnarray}
& & |\Delta \omega| < \gamma < \frac{1}{3}(2 + \sqrt{1-3\Delta \omega^2}),
\label{eq:range_rocking}
\end{eqnarray} 
as follows from the analysis of relations (\ref{eq:solutions}). 
In particular the condition $\sqrt{3}\Delta \omega < 1$ is required. The
analytical estimate of the locking range given by Eq.~(\ref{eq:range_rocking})
is plotted as solid lines in
Figs.~\ref{fig:thlock} and \ref{fig:f04_theo}.
The entrainment range has been evaluated considering only one harmonic
component of the driving signal (the one closest to resonance), and
solving Eq.~(\ref{eq:landau}) with $A_{\rm inj}(\tau)=A_0 \exp(i\Omega\tau)$.
The locking boundary is then given by $A_0=\mid\Omega-\Delta\omega\mid$,
as a simple analysis of Eq.~(\ref{eq:landau}) shows. For a step-like
injection $A_{\rm inj}=A_0\; \mathrm{sign}[\sin(\Omega \tau)]$, the rocking
condition reads
\begin{equation}
A_0=\mathrm{mod}(\Omega-\Delta\omega)\frac{\pi}{2}\,.
\label{eq:lrstep}
\end{equation}

A physical interpretation of both the lower and upper rocking conditions in
Eq.~(\ref{eq:range_rocking}) is as follows.
For relatively large amplitudes of rocking, $A_0 > A_0^{\rm max}(\Omega)$, the
slow component of the oscillations $a(t)$ decays to zero and one is left with the
fast oscillating component of the amplitude. Therefore, in that case the amplitude
of the oscillations follows adiabatically the oscillations of the driving force. On the
other hand, for small values of the 
rocking amplitude, $A_0<A_0^{\rm min}(\Omega)$, the phase of the oscillator is unlocked
from that of the driving force, and evolves freely.

We integrated numerically, by means of a second order Runge-Kutta scheme,
the complex Landau equation (\ref{eq:landau}), in order to verify
the evaluated rocking ranges. The rocking area was indeed found to
decrease with increasing detuning $\Delta\omega$, as depicted by the white
circles in Fig.~\ref{fig:thlock}. Increasing the frequency of rocking $\Omega$ also
leads to an increase of the rocking range, as depicted in \ref{fig:f04_theo}.
Both behaviors are in qualitative and quantitative agreement with the analytical result
given by Eq.~(\ref{eq:range_rocking}). 

\begin{figure}[htb]
\centering
\includegraphics[width=70mm,clip]{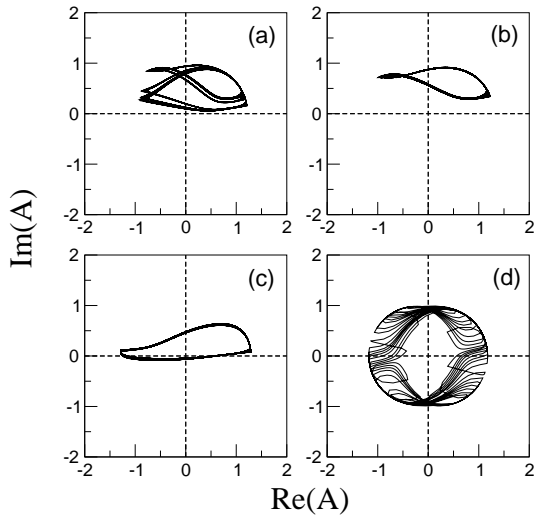}
\centering
\caption[2]{
Trajectories in phase space for the situations depicted as black circles in
Fig.~\protect\ref{fig:f04_theo} above. Plots (a-c) correspond to the rocking
regime: (a) lies on the boundary with the free-running regime, (b) lies deep within the
rocking region, and (c) is located on the boundary with the adiabatic regime.
(d) lies outside the rocking regime.
Parameter values are $\Delta\omega=0.1$ and:
(a) $\Omega=0.7$, $A_0=0.61$,
(b) $\Omega=1.0$, $A_0=0.85$,
(c) $\Omega=0.7$, $A_0=0.87$, and
(d) $\Omega=0.55$, $A_0=0.45$.
} 
\label{fig:f05_theo}
\end{figure}

The different dynamical regimes observed in our numerical calculations are 
summarized in 
Fig.~\ref{fig:f05_theo}, which shows typical phase trajectories on the complex plane 
of the oscillation amplitude. Locking corresponds to 
closed trajectories, which are placed symmetrically with respect to 
the Re$(A)$-axis. 
Besides the periodic regimes displayed in Figs.~\ref{fig:f05_theo}(b,c),
more complicated ones were also 
found in the locking area (predominantly at its border), 
such as period doubled and chaotic orbits. Examples of these 
behaviors are given in plots (a) and (d) of Fig.~\ref{fig:f05_theo}.   
   
\section{Conclusions}

We have provided for the first time experimental evidence of the phenomenon
known as rocking in a nonlinear electronic oscillator. Rocking allows the control
of the oscillation phase, which furthermore takes the form of a bistable
response: in the presence of a harmonic driving whose phase is
periodically inverted in time (at a much smaller frequency than that of
the carrier oscillations), the system responds with one of two phases
orthogonal to the input phase. Coexistence of these two phases 
can be revealed by multiple entrances into the locking region, and can occur
spontaneously in the boundaries.

The basic properties of the rocking (phase bistability and transition to the
adiabatic and free-running regimes) can be well understood in terms of normal
form analysis, i.e. by investigating the Òrocked Hopf bifurcationÓ by means of
the complex Landau equation description (\ref{eq:landau})-(\ref{eq:range_rocking}).
There are some quantitative discrepancies, especially at large amplitudes of the
external drive, when the normal form description (valid close to threshold)
no more holds. Also it seems (see Figs.~\ref{fig:f03} and \ref{fig:f05}) that the
Chua oscillator displays a weak nonlinear resonance (i.e. the resonance frequency
depends on the oscillation amplitude) that is not considered in the normal form
description (\ref{eq:landau}). Those specific aspects of rocking could be observed
by modelling the full equations (\ref{eq:chuatot})-(\ref{eq:piece}) of the circuit. 

These results
could provide a means to encode information in phase-based communication
systems.

\bigskip
\section*{Acknowledgments}

Discussions with Germ\'an J. de Valc\'arcel are acknowledged.
This work was financially supported by the Ministerio de Educaci\'on y Ciencia
(Spain) through projects BFM2003-07850, FIS2005-07931-C03-03, and
TEC2005-07799, and by the Generalitat de Catalunya.

\end{document}